# A model for cage formation in colloidal suspension of laponite

Yogesh M. Joshi

Department of Chemical Engineering, Indian Institute of Technology Kanpur, Kanpur 208016, INDIA.

E-Mail: joshi@iitk.ac.in

Telephone: +91 512 259 7993, Fax: +91 512 259 0104

**Abstract**

In this paper we investigate glass transition in aqueous suspension of synthetic hectorite clay, laponite. We believe that upon dispersing laponite clay in water, system comprises of clusters (agglomerates) of laponite dispersed in the same. Subsequent osmotic swelling of these clusters leads to increase in their volume fraction. We propose that this phenomenon is responsible for slowing down of the overall dynamics of the system. As clusters fill up the space, system undergoes glass transition. Along with the mode coupling theory, proposed mechanism rightly captures various characteristic features of the system in the ergodic regime as it approaches glass transition.

**INTRODUCTION:**

Glasses are disordered materials, and in such state, system explores only a small part of the phase space. Molecular glasses are formed by quenching the liquid rapidly below so called glass transition temperature to obtain a disordered structure.[1] In colloidal suspensions, the glassy state is attained by increasing the concentration of the constituent particles such that the disordered state is obtained above a random loose packing threshold.[2] Generally this is achieved by either centrifugation[3] or dialysis.[4] Thus, colloidal glasses are different from molecular glasses wherein the concentration, not the temperature, plays an important role in causing arrested or glassy state. In this paper we investigate the kinetics of glass transition in colloidal glass of aqueous laponite suspension wherein osmotic swelling of laponite clusters brings about glass transition.

Lately, the colloidal glasses formed by aqueous suspensions of laponite have developed considerable interest.[5-10] Laponite particle has a disc-like shape, and due to negative charges on its surface, repulsive interactions prevail among particles in low ionic concentration aqueous medium that leads to a formation of the so called Wigner glass.[10] Apart from having wide ranging industrial applications; these suspensions are considered as model systems to study slow dynamics of glass transition and gelation.[11]

The laponite disc has diameter of 25 nm and layer thickness of 1 nm. At very low ionic concentrations its suspension in ultra pure water leads to a glassy state for volume fractions less than 0.01 in contrast to colloidal glasses of uncharged spherical particles where glassy state is attained at volume fractions only above 0.5. This is essentially due to anisotropic shape of a laponite disc and negatively charged surface. Later gives rise to electrostatic screening length of 30 nm.[12] Hence the effective volume of a single laponite disc increases by a factor of 60 and explains the requirement of approximately 60 times lower volume fraction for glass formation. (Volume of single laponite particle is of the order of $1\times25\times25$ nm$^3$. However the effective volume of laponite particle including electrostatic screening length of 30 nm on both the sides becomes $60\times25\times25$ nm$^3$)

In the past few years several groups have studied liquid-glass transition of a laponite suspension at low ionic concentration using various optical techniques.[5-10,12-14] In general laponite powder is mixed in ultra-pure water with vigorous stirring and passed through a micro-filter to study the evolution of its structure with respect to age. The suspension shows two relaxation modes with the fast mode being independent of age.[13,14] Its inverse square dependence on wave vector ($q$) highlights its diffusive character. The slow mode shows an initial rapid increase followed by a linear increase with respect to age.[5,15] The former regime is called the cage forming regime,[15] where the slow mode also shows inverse square dependence on $q$. In the full aging regime, it shows $q$ dependence with power law index in the range –1 to –1.3. Furthermore, the age at which transition from rapid growth to linear growth occurs, decreases with increase in clay concentration.[5] Very recently Mossa et al.,[16] while studying ageing of laponite suspension in the nonergodic regime using Brownian dynamics simulations, observed that anisotropy of the platelet along with repulsive

Yukawa potential is responsible for the local disorder that takes the system into the metastable state. They further observed that ageing dynamics strongly affects orientational degree of freedom which relaxes over the timescale of translational modes. There are various other morphologies proposed in the literature, and in general the state of the system in full ageing regime, gel or glass, still remains a contentious issue.[17]

In the literature, the rapid growth of relaxation time with respect to age in the ergodic regime is fitted by an empirical function $\left(\ln \tau \sim t_w\right)$ and hence is generally referred to as an exponential growth.[5,13-15] Tanaka *et al.*[15] proposed two-stage aging kinetics by expressing the relaxation time by using the average barrier height $U$ for the particle motion as: $\tau \sim \exp(U/kT)$. They suggested that the barrier height should grow linearly with aging time in the ergodic regime while logarithmically in the full aging regime to yield exponential and linear age dependence of relaxation time respectively. However they mentioned that the basis for logarithmic dependence that leads to exponential growth is not clear. They concluded by asserting that that rapid increase in slow mode with age in the ergodic regime remains an unanswered question. Schosseler *et al.*,[5] after experimentally investigating this phenomenon, linked it to liquid-glass transition, but did not conclude about the mechanism. In this paper we investigate this very phenomenon. For the first time, we show that the osmotic swelling of laponite clusters bring about rapid increase in the relaxation time as the system approaches arrested state. We limit our discussion only to cage forming regime.

**THEORY:**

We consider a system of aqueous suspension of laponite having basic pH that leads to net negative charge on the laponite particle.[18] The chemical formula for laponite is $Na^+_{0.7}[(Si_8Mg_{5.5}Li_{0.3})O_{20}(OH)_4]^-_{0.7}$. Isomorphic substitution of magnesium by lithium atoms generates negative charges on its surface that are counterbalanced by the positive charge of the sodium ions present in the interlayer. We assume that immediately after stirring/filtration the system comprises of clusters (agglomerates) of laponite dispersed in water. In the present model, the word cluster should be referred to as agglomerate that is composed of domains of parallel stacks of laponite particles.

As time progresses, the clusters of laponite undergo swelling due to diffusion (flux) of water into the cluster driven by an osmotic pressure gradient that is caused by repulsive interactions between neighboring particles.

We believe that the clusters of laponite get hydrated while preparing the suspension by stirring. Filtration process breaks the clusters. For simplicity we consider all the clusters to be spherical in shape, of same size and at the same level of hydration. Let $\psi$ be the volume fraction of laponite inside a cluster of radius $R_0$ at time $t_w=0$. Just outside the cluster, the volume fraction of water is unity and imposes an osmotic pressure gradient. As time progresses water diffuses into the cluster until the concentration of water (or alternatively laponite, or the osmotic pressure) becomes uniform over the space. If the overall volume fraction of laponite is $\phi$, the number of clusters per unit volume is given by,

$$n = \phi/(\psi R_0^3). \tag{1}$$

The clusters swell due to osmotic diffusion. If the $R(t_w)$ is the radius of cluster at any time $t_w$, then the fraction of total volume occupied by clusters is $\xi(t_w)^3 \phi/\psi$ while the mode coupling distance parameter $\varepsilon$ is given by:

$$\varepsilon = \left[\phi_c - \xi(t_w)^3 \phi/\psi\right]/\phi_c, \tag{2}$$

where $\xi(t_w) = R(t_w)/R_0$ is the dimensionless radius and $\phi_c$ (~0.56-0.64) is the volume fraction at which growing spherical clusters touch each other. We believe that as $\varepsilon \to 0$; the system undergoes a transition from a cage forming regime to a glassy or full aging regime. The dimensionless radius at the onset of this transition can be obtained from eq. (2) and is given by,

$$\xi^* = (\phi_c \psi/\phi)^{1/3}. \tag{3}$$

In the powder form, laponite discs are present in the domains of parallel stacks with sodium atoms present in the interlayer. Upon dispersing in water, swelling of clay occurs in two steps. In the first step, water is absorbed in successive monolayers on the surface that pushes the discs apart. As the hydration of clay takes place, the Na$^+$ ions, though are electrostatically attracted towards the oppositely charged surface, diffuse away from the surface into the bulk where their concentration is low. As some Na$^+$ ions diffuse out of the gallery, the negatively charged surfaces of the opposite faces get exposed to each other and hence repel. In the second stage of swelling, double layer repulsion pushes the laponite discs further apart.[18] The larger the distance between the laponite discs, lower is the osmotic pressure required to hold them in the same position. Recently Martin and coworkers[19] proposed a scaling model to estimate flux of water through concentrated sediment of laponite under an osmotic pressure gradient. Using analytical solution of Poisson-Boltzmann equation for the case where only counterions are present in the interlayer space,[20] they argued that the osmotic pressure varies as the inverse square of the plate separation distance in the concentration regime of usual interest [namely, plate separation distance 0.5 nm to 30 nm for the present system that corresponds to concentration of about 33 vol % to 1.6 vol %]. Their expression for dependence of osmotic pressure on clay concentration gives excellent prediction of the experimental data. Since the osmotic pressure gradient drives the flow of water through porous sediment, they used Darcy's law to predict flow rate. If $\Delta P$ is the pressure difference over the radius of the cluster, the flux of water having viscosity $\mu$ is given by $Q = k_p A \Delta P / (\mu R)$, where $k_p$ is a dimensionless constant that depends on the porosity and characteristic features of the cluster (porous object) and $A$ is the surface area of the sphere. Considering an incremental swelling step, where flux $Q$ over time $dt_w$ causes an increase in volume by $dV$, we get $Q dt_w = dV$.[19] This leads to $\mu R dR = \Delta P k_p dt_w$. Based on the analytical solution for Poisson-Boltzmann equation, Martin and coworkers[19] argued that dependence of product $k_p \Delta P$ on volume fraction of laponite in the cluster (and hence time) can be neglected in the concentration regime of interest. The above equation can be easily integrated to predict the size of the swelling cluster under osmotic pressure gradient with respect to time to yield,

$$\xi(t_w) = \left[1 + \left(D t_w / R_0^2\right)\right]^{1/2}. \tag{4}$$

where $t_w$ is waiting time and $D$ is characteristic diffusivity of the swelling process in the limit of small volume fraction of laponite. It is given as, $D = \dfrac{2\Delta P k_p}{\mu} \approx \dfrac{4(\pi k T)^2 \varepsilon_0 \varepsilon_r a^2}{45 e^2 t_l^2 \mu}$, where $k$ is Boltzmann constant, $T$ is absolute temperature, $\varepsilon_0$ is permittivity of vacuum, $\varepsilon_r$ is relative permittivity of water medium, $e$ is unit charge, $t_l$ is thickness of laponite plate and $a$ is equivalent radius that gives the same mass per particle as a laponite particle (0.82 nm).[19] Thus the characteristic diffusivity can be readily evaluated by knowing the temperature of the system.

Immediately after filtration, the system is filled with many clusters undergoing Brownian motion. As the volume fraction of clusters, $\xi(t_w)^3 \phi / \psi$ goes on increasing, the available space that is not occupied by clusters decreases and the dynamics of the system slows down. This increases the corresponding characteristic time-scale of the system. This process is akin to glass transition, where, as temperature of a glass forming liquid is decreased, movement of the constituent molecules gets constrained. If we apply this analogy to the present system, the growth of the cluster can be considered equivalent to decreasing the temperature of the molecular glass formers. Under such conditions we can estimate the characteristic relaxation time of the system using mode coupling theory that describes the behavior of colloidal glasses very well. The relaxation time predicted by the mode-coupling theory is given by:[21]

$$\tau = \tau_0 \varepsilon^{-\gamma}, \tag{5}$$

where, $\tau_0$ and $\gamma$ are the fitting parameters.

The characteristic relaxation time can be obtained by incorporating eqs. (2) and (4) into eq. (5). The model has three fitting parameters, volume of laponite in the cluster $(R_0^3 \psi)$, $\tau_0$ and $\gamma$. The first parameter arises naturally in the formulation and cannot be prescribed a priori. This is because the radius of cluster and the volume fraction of laponite inside a cluster at time $t_w = 0$ are strongly dependent on the time of stirring. The latter parameter $\tau_0$ is observed to be of order unity for the present system. Eq. (4) can be incorporated in eq. (3) to predict the transition time at which system enters a nonergodic or full aging regime and is given by:

$$t_w^* = \phi_c^{2/3} \left(R_0 \psi^{1/3}\right)^2 / \left(D \phi^{2/3}\right). \tag{6}$$

However, due to uncertainty in getting reproducible data,[5] it is difficult to validate this expression experimentally as it demands $(R_0^3 \psi)$ to be identical in every experiment. Combining eqs (2) to (6), we get:

$$\tau/\tau_0 = \left[1 - \left(t_w/t_w^*\right)^{3/2}\right]^{-\gamma} \quad \ldots \quad t_w < t_w^*. \tag{7}$$

Thus, knowledge of $\tau_0$ and $t_w^*$ determine nature of evolution of relaxation time for given value of $\gamma$.

**DISCUSSION:**

In figure 1, the characteristic dimensionless time scale $(\tau/\tau_0)$ of the suspension is plotted as a function of dimensionless age of the sample $(t_w/t_w^*)$ for 2 and 2.75 wt. % suspension of laponite in water,[5] while an inset shows a log-log plot of $\tau/\tau_0$ vs. $1-(t_w/t_w^*)^{3/2}$. As shown in the figure, eq. (7) with $\gamma=2.58$ provides an excellent fit to the experimental data. Since the experimental data is made dimensionless by the model parameters that are obtained from the experiments (namely $\tau_0$ and $t_w^*$), only a unique value of $\gamma$ can fit the given data set. Interestingly $\gamma=2.58$ is the same value for which classic mode coupling theory predicts glass transition in the monodispersed spherical particles.[21] However, the real sample of laponite suspension is expected have size distribution of clusters and hence a fit of $\gamma=2.58$ to the experimental data might be coincidental. As the volume fraction of the clusters approaches $\phi_c$, the characteristic time-scale approaches the age of the system and undergoes a transition from a cage forming regime (ergodic) to a full aging regime (nonergodic). This might not be a sharp transition as the glassy state is attained only after laponite particles occupy the available space completely, thus highlighting role of the length scale probed in the experiment. Consequences of this result are discussed later in the paper. In the full aging regime, system shows a linear relationship between characteristic timescale and age. In the present model, in order to keep analysis simple, we have considered osmotic swelling of clusters having same initial radius. According to eq. (4), consideration of polydispersity in the initial radius of cluster will lead to change in polydispersity with age. We believe that simplistic model and assumptions proposed in the present manuscript is the first step to theoretically investigate the ergodicity breaking mechanism in this system. A quantitative prediction provided by the present model is indeed encouraging in that respect.

The relaxation time obtained by eq. (7) is a slow mode representing the characteristic time associated with the cage diffusion process. In many experiments that observe the rapid increase in slow relaxation mode with respect to time employ sub- micron size tracer particles to strengthen the signal.[5] As clusters grow, movement of these tracer particles gets confined to smaller and smaller region in space. Although this does not significantly affect characteristic time associated with rattling motion within the confined space (fast mode), time required to escape the cage formed by neighboring clusters increases rapidly. However the dynamics of the tracers is still diffusive in nature leading to $q^{-2}$ dependence. As growing laponite clusters fill the available space completely, cage diffusion becomes extremely sluggish and the corresponding hyperdiffusive timescale scales as age.

The concept modeled in this paper is similar to one of the ideas proposed by Tanaka *et al.*[15] where they speculate that the system becomes nonergodic after the growing clusters of aggregates fill up the space. However, the present model is very different than that proposed by the same group,[15] where they argue that the average barrier height for particle motion grows linearly with age in the ergodic regime. They estimate the conductivity of the laponite suspension with respect to its age. They found that the conductivity increases almost linearly with the aging time in the ergodic regime, but tends to saturate in the nonergodic regime. The increase in conductivity reflects the reduction in the number of strongly bound counterions. Our model is in agreement with this observation. As the cluster size increases, more and more counterions diffuse away into the bulk increasing the conductivity. As the clusters fill up the space, along with laponite particles, the concentration of counterions also becomes uniform, leading to saturation in the conductivity.

The present model clearly distinguishes between the ergodic state and non-ergodic state based on the physical structure that exists in these two states. In the former state clusters or agglomerates of laponite platelets are present, while in the later regime, single laponite particle is an independent entity. It is well known that aqueous suspension of laponite, when it is in the nonergodic regime, undergoes rejuvenation due to excessive deformation.[22] However, according to proposed physical picture, the model clearly states that, due to rejuvenation, the system cannot cross the ergodic-nonergodic transition point and enter ergodic regime. This means that exponential-like rapid increase in slow relaxation mode cannot be observed again. This observation has significant implications in analyzing rheological behavior of laponite suspensions. Furthermore eq. (6) predicts inverse relationship between transition time and characteristic diffusivity. We have seen that temperature dependence of characteristic diffusivity $\left(D \propto (kT)^2 \varepsilon_r / \mu \right)$ comes from three terms, namely, $(kT)^2$, permittivity of water and viscosity of water. Viscosity of water can be considered to depend on temperature as $\mu = \mu_0 e^{\bar{U}/kT}$, while permittivity of water, though explicit analytical expression for its dependence on temperature does not exist,[23] decreases weakly compared to that of viscosity of water. Overall, transition time is expected to show significant decrease with respect to temperature. Ramsay[24] studied effect of temperature on rheological properties of ageing laponite suspension and observed pronounced increase in elastic modulus with age at higher temperature. This observation matches very well with the prediction of the model.

Nicolai *et al.*[25] carried out static light scattering experiments on the aqueous laponite dispersions in the concentration range 0.025 wt % to 0.5 wt. %. They observe that intensity of scattered light decays significantly in the initial five hours followed by a very sluggish decay. Dynamic light scattering experiments on the same sample after one day showed system to be still ergodic, as expected for such low concentration of laponite and suggested presence of individual laponite particles or an incomplete dispersion of the oligomers with a broad size distribution. An initial decay in the intensity of scattered light that eventually leads to oligomers or individual laponite particles can be very well explained by the present model. As various clusters grow in size, water content in the same increases, which decreases relative difference in the refractive index between water and the cluster, decreasing the intensity of scattered light.

Schosseler *et al.*[5] have discussed various features of ergodic to nonergodic transition in great details. They observed that immediately after filtration, the viscosity of the suspension as recorded by the diffusivity of the tracer particles is of the order of few mPas. This observation further strengthens the assumption that laponite is present in the form of tiny clusters immediately after filtration. Schosseler *et al.*[5] further observed that that the full aging behavior is first seen while investigating large length scales in the aging suspension of laponite. In the present paper we argue that transition to nonergodic regime occurs when growing clusters of laponite touch each other. However, the space between the clusters when they touch each other still contains low viscosity aqueous medium which is ergodic in nature. Thus, the transition to full ageing regime will not be observable until the probed length scale is larger than the space between the growing clusters. Model captures this behavior very well. Thus proposed model rightly captures various experimental observations in the ergodic regime of aqueous suspension of laponite.

**CONCLUSION:**

We have modeled a new mode of glass transition in which clusters of laponite particles undergo osmotic swelling and enter the non-ergodic state as they span the available space. As the clusters fill up the space, cage diffusion process becomes very sluggish. The mode coupling formalism along with proposed mechanism provides an excellent prediction of the associated relaxation time dependence on age. Model also predicts that the ergodic to nonergodic transition is first observed at large length scales and it occurs at early age for higher temperature. These predictions are in agreement with experimental observations.

**ACKNOWLEDGEMENT:**

Financial support from Department of Atomic Energy, Government of India under the BRNS young scientist award scheme is greatly acknowledged. I would like to thank Dr. Ranjini Bandyopadhyay, Dr. S. A. Ramakrishna and Dr. G. Kumaraswamy for constructive remarks and discussion.

~~~~~~~~~~

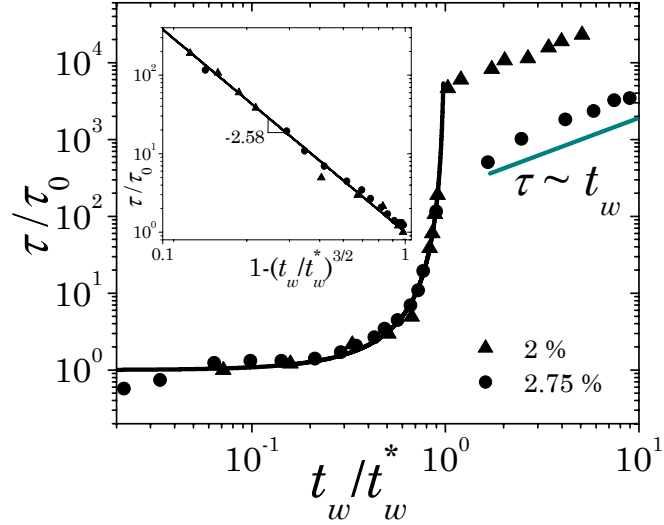

**Figure 1.** The characteristic dimensionless time scale $(\tau/\tau_0)$, of the suspension is plotted against dimensionless age of the sample $(t_w/t_w^*)$ for 2 and 2.75 wt. % suspension of laponite in water. Line is eq. 7 while the experimental data is taken from Figure 2 of Schosseler *et al.*[5] For 2 % sample, $\tau_0 = 2.9$ s and $t_w^* = 13600$ s while for 2.75 % sample $\tau_0 = 0.94$ s and $t_w^* = 5500$ s. Inset shows the same data and the fit, plotted against $1-(t_w/t_w^*)^{3/2}$, in the ergodic regime. A power law with exponent –2.58 uniquely fits the data.